\documentclass[useAMS]{emulateapj}
\usepackage[]{natbib}
\usepackage{graphicx}

\def\lsim{\lower.5ex\hbox{$\; \buildrel < \over \sim \;$}}
\def\gsim{\lower.5ex\hbox{$\; \buildrel > \over \sim \;$}}

\shorttitle{Detection of CRSF in 4U~1909+07}
\shortauthors{Jaisawal et al.}

\begin{document}
\title{Possible detection of a cyclotron resonance scattering feature in the X-ray pulsar 4U~1909+07}

\author{Gaurava K. Jaisawal$^{1}$, Sachindra Naik\altaffilmark{1}, Biswajit Paul\altaffilmark{2}}

\altaffiltext{1}{Astronomy \& Astrophysics Division, Physical Research Laboratory, Ahmedabad - 380009, India. $gaurava@prl.res.in$}
\altaffiltext{2}{Raman Research Institute, Sadashivnagar, C. V. Raman Avenue, Bangalore 560080, India}

\begin{abstract}
We present timing and broad-band spectral studies of the high mass X-ray binary pulsar 
4U~1909+07 using data from $Suzaku$ observation during 2010 November 2-3. The pulse 
period of the pulsar is estimated to be 604.11$\pm$0.14 s. Pulsations are seen in the 
X-ray light curve up to $\sim$70 keV. The pulse profile is found to be strongly 
energy-dependent: a complex, multi-peaked structure at low energy that becomes a 
simple single peak at higher energy. We found that the 1-70 keV pulse averaged 
continuum can be fitted by the sum of a black body and a partial covering Negative 
and Positive power-law with EXponential cutoff (NPEX) model. A weak iron 
fluorescence emission line at 6.4 keV was detected in the spectrum. An 
absorption like feature at $\sim$44 keV was clearly seen in the residue of the 
spectral fitting, independent of the continuum model adopted. To check the 
possible presence of a CRSF in the spectrum, we normalized the pulsar spectrum 
with the spectrum of the Crab Nebula. The resulting Crab ratio also showed a clear 
dip centered at $\sim$44 keV. We performed statistical tests on the residue of 
the spectral fitting and also on the Crab spectral ratio to determine 
the significance of the absorption like feature and identified it as a CRSF of 
the pulsar. We estimated the corresponding surface magnetic field of the pulsar 
to be 3.8$\times$10$^{12}$ Gauss.

\end{abstract}

\keywords{X-rays: stars -- neutron, pulsars -- stars: individual -- 4U~1909+07}

\section{Introduction}

Accretion powered X-ray binary pulsars which were discovered in the early 70s 
(Giacconi et al. 1971), are among the brightest X-ray sources in sky. These 
binary systems consist of a neutron star with strong magnetic field ($B \sim 
10^{12}$ Gauss) and a supergiant or a Be star as optical companion. 
Mass transfer from the companion star to the neutron star takes place through 
Roche lobe overflow (in case of low mass X-ray binaries such as 
Her~X-1, Reynolds et al. 1997) and/or capture of stellar wind (in high mass 
X-ray binary systems). The accreted matter is channeled towards the magnetic 
poles of the neutron star by its strong magnetic field forming accretion columns 
from where the gravitational energy of accreted matter is being 
dissipated in the form of X-rays. Most of accretion powered X-ray pulsars 
belong to the group of high-mass X-ray binary (HMXB) systems. Based on spectral 
and luminosity class of companion stars, the HMXBs are further classified into 
two subgroups such as (i) Supergiant X-ray binary (SGXB) systems and (ii) 
Be/X-ray binary systems. In case of SGXBs, the optical companion is 
an OB supergiant star whereas the Be/X-ray binary systems associate with non-supergiant 
B-type stars which show Balmer emission lines in their spectra. 

The energy spectra of accretion powered X-ray pulsars are generally described by
phenomenological models consisting a power-law of photon index $\sim$1, high energy 
cut-off (dal~Fiume et al. 1998) and a Gaussian function at 6.4 keV for the presence of iron
fluorescence emission. In some cases, a black-body or bremsstrahlung component 
is required to describe the presence of excess emission at soft X-rays (Paul et al. 
2002; Naik \& Paul 2004a, 2004b; Hickox, Narayan \& Kallman 2004). The broad-band 
X-ray spectrum of some pulsars has been described with NPEX continuum model 
which is an approximation of the unsaturated thermal Comptonization in hot plasma 
(Makishima et al. 1999). The NPEX continuum model reduces to a simple power-law with 
negative slope at low energies that is often used to describe the spectra of accretion 
powered X-ray pulsars. In case of transient Be/X-ray binary pulsars, these continuum 
models are marginally modified to describe the presence of absorption features at 
certain pulse phases. Partial covering high energy cut-off power-law model is being 
used to describe the broad-band spectrum of many transient pulsars (Paul \& Naik 2011;
Naik et al. 2011 and references therein). This model consists of two power-law continua 
with a common photon index but with different hydrogen-absorbing column densities. 
Several emission lines due to fluorescence from ions at different ionization levels 
and broad absorption like features due to the cyclotron resonance scattering are 
often seen in the pulsar spectrum. The magnetic field strength $B$ and the 
cyclotron resonance energy $E_a$ are related through the relation E$_a=11.6 B_{12} 
(1+z_g)^{-1}$ (keV), where $z_g$ is gravitational red-shift and B$_{12}$ is magnetic 
field strength in units of 10$^{12}$ Gauss. Detection of CRSF in the spectrum, therefore, 
provides the direct measurement of the strength of the pulsar magnetic field. The CRSFs 
have been detected in the spectrum of about 19 X-ray pulsars (Coburn et al. 2002; 
Staubert et al. 2003; Pottschmidt et al. 2012) using data from several X-ray observatories. 
The harmonics of fundamental cyclotron absorption line are also detected in some pulsars 
(Nakajima et al. 2006; Orlandini et al. 2012). 
	
	The HMXB pulsar 4U~1909+07 was discovered with  $Uhuru$ satellite and mentioned 
as 3U~1912+07 in  3rd $Uhuru$ catalogue (Giacconi et al. 1974). The position of the
pulsar was refined and the source was renamed as 4U~1909+07 in 4th catalogue (Forman 
et al. 1978). The presence of many other sources in the intensity range of 2-12 mCrab
and 2-10 keV energy range was reported at nearby co-ordinates of 4U~1909+07 through 
observations with different missions such as $OSO~7$, $Ariel~5$, $HEAO-1$, $EXOSAT$.
Wen et al. (2000) recognized that all the sources such as 4U~1909+07, 3A~1907+074, 
1H~1907+074, GPS~1908+075, 1E~1908.4+0730 and X1908+075 reported through various surveys 
are same in uncertainty of $1'$. The orbital period of the binary system was reported 
to be 4.4 days using $Rossi~X-Ray~Timing~Explorer (RXTE)$ All-Sky Monitor (ASM) data 
(Wen et al. 2000). Using $RXTE$ Proportional Counter Array (PCA) observations of
4U~1909+07, pulsations of 605 s were detected in the X-ray flux (Levine et al. 2004).  
The orbital inclination, orbital separation and mass of the companion star were estimated 
to be in the ranges of 38${^\circ}$--72$^\circ$, 60--80 lt-s and 9--31 M$_\odot$, 
respectively (Levine et al. 2004). The detection of an OB star in near-infrared within 
the X-ray error box of the pulsar confirmed the system to be a OB supergiant-neutron 
star HMXB (Morel \& Grosdidier 2005). The distance of the binary system was 
estimated to be $\sim$7 kpc (Morel \& Grosdidier 2005). $RXTE$ and $INTEGRAL$ observations 
of the pulsar showed that 605 s pulsation in the X-ray flux is not stable, rather it 
changes erratically on time scales of years (F{\"u}rst et al. 2011). The pulse profile 
was found to be strongly energy dependent. The phase averaged spectrum, obtained from 
above observations, were well described by a power-law with high energy 
cut-off continuum model along with a black-body component (F{\"u}rst et al. 2011).
F{\"u}rst et al. (2012) also analyzed data from  $Suzaku$ observation of the pulsar and 
described the spectrum using same model as was used in case of their earlier work. In both
cases, there was no detection of CRSF in the pulsar spectrum. Data from $Suzaku$ 
observation was used in  later case, the high energy data was truncated at 40 keV 
in the spectral fitting. However, using the same dataset up to high energy ranges, 
we detected an absorption like feature at $\sim$44 keV and interpret as possible 
CRSF in the pulsar spectrum. The details of analysis and results obtain are 
described in the following sections.

\begin{figure}
\centering
\includegraphics[height=3.3in, width=3.05in, angle=-90]{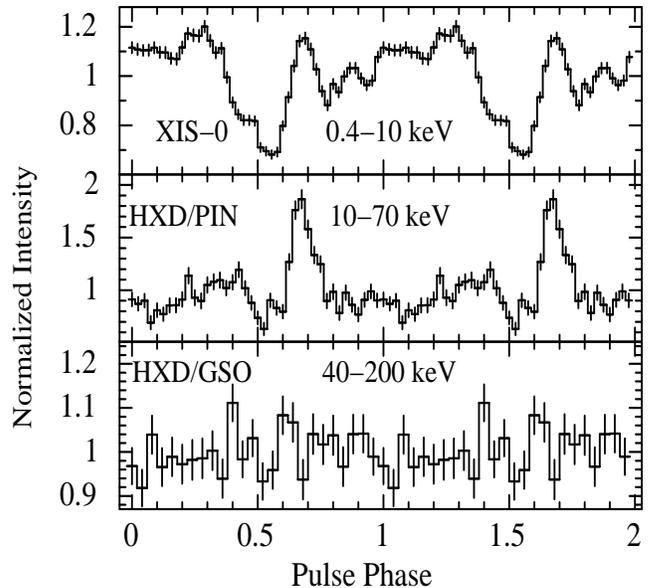}
\caption{Pulse profiles of the HMXB pulsar 4U~1909+07 in 0.4-12 keV (XIS-0), 10-70 keV (HXD/PIN) 
and 40-200 keV (HXD/GSO) energy ranges are shown. The profiles are obtained from corresponding
light curves by estimated 604.11 s pulse period of the pulsar. The errors shown in the figure 
are estimated for 1$\sigma$ confidence level. Two pulses are shown for clarity.}
\label{pp}
\end{figure}

\section{Observation and Analysis}

The HMXB pulsar 4U~1909+07 was observed with $Suzaku$ on 2010 November 2-3.   
We used the publicly available archival data of processing version 2.5.16.28
in present work to investigate timing and spectral properties of the pulsar. 
The observation was carried out at ``XIS nominal'' pointing position for effective
exposures of $\sim$30 ks and $\sim$22 ks for X-ray Imaging Spectrometers (XIS) and 
Hard X-ray Detectors (HXD), respectively. XIS were operated with ``normal'' 
clock mode in ``1/4 window'' option. In this operation mode, the time resolution 
and field of view of XIS are 2 s and 17$'$.8$\times$4$'$.4, respectively.

$Suzaku$, the fifth Japanese X-ray astronomy satellite, was launched by Japan Aerospace 
Exploration Agency (JAXA) on 2005 July 10 (Mitsuda et al. 2007). The instruments onboard
$Suzaku$ covers 0.2-600 keV energy range through two sets of instruments, XIS (Koyama et 
al. 2007) and HXD (Takahashi et al. 2007). XISs are imaging CCD cameras which are 
located at the focal plane of X-ray telescopes (XRTs). Among the four XISs, XIS-0, XIS-2 
and XIS-3 are front illuminated (FI) whereas XIS-1 is back illuminated (BI). In full window 
mode, the effective area of XIS is 340 cm$^2$ for FI and 390 cm$^2$ for BI at 1.5 keV. Due 
to large charge leakage in the imaging region, XIS-2 is no more operational since 2007 
September. The non-imaging detector HXD consists of two types of instruments such as silicon 
PIN diodes covering 10--70 keV energy range and GSO crystal scintillator covering 30--600 keV 
energy range. The effective area of PIN is 145 cm$^2$ at 15 kev and for GSO is 315 cm$^2$ at 
100 keV.  Field of view (FoV) of XIS and PIN are $18'\times18'$ and $34'\times34'$ in open 
window mode, respectively. GSO has  same FoV as PIN up to 100 keV. As XIS-2 is no more 
operational, data from other 3 XISs, PIN and GSO are used in the present analysis.

\begin{figure*}
\centering
\includegraphics[height=7.in, width=4.in, angle=-90]{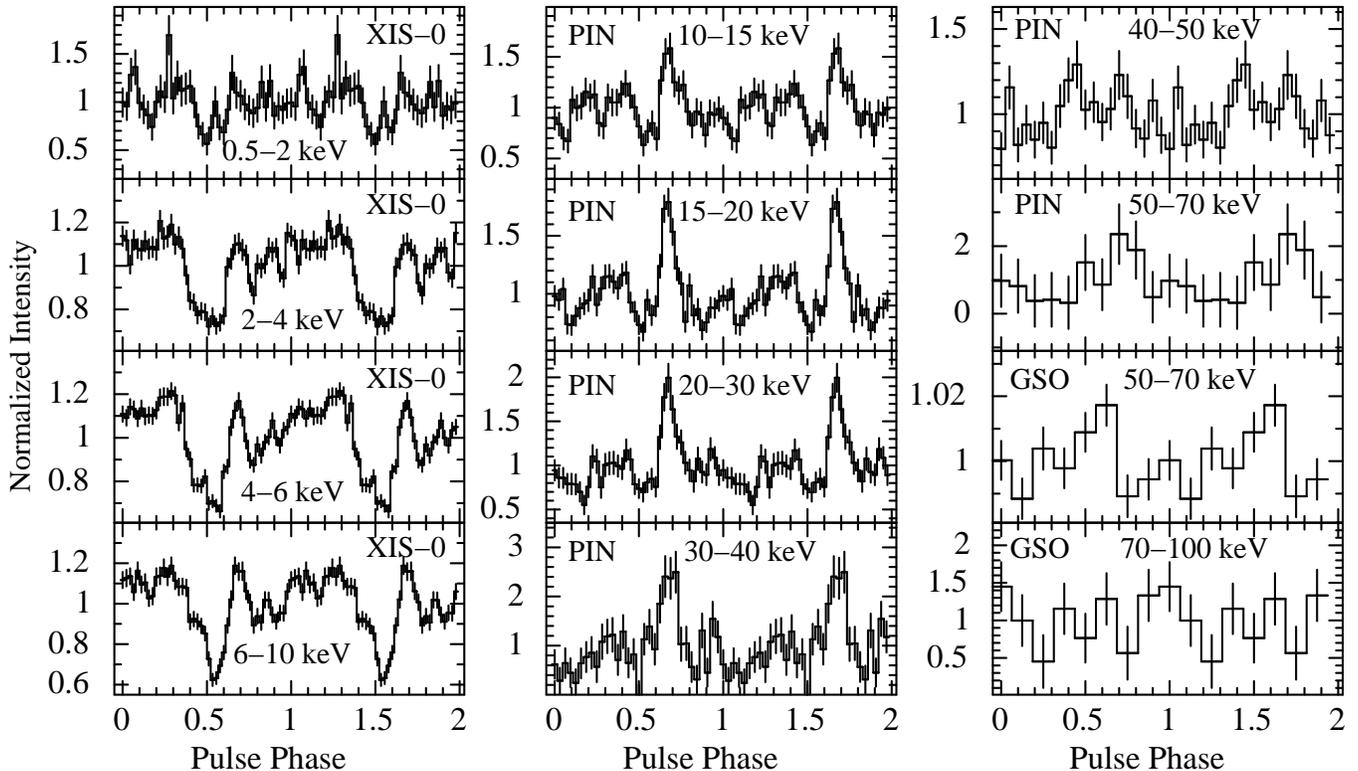}
\caption{The energy resolve pulse profile of 4U~1909+07 at different energy bands obtained
from XIS-0, HXD/PIN and HXD/GSO data. The evolution of pulse profile from complex shape in soft
X-ray energy ranges to a single-peaked profile up to $\sim$40 keV can be clearly seen. The 
pulsation in GSO can be seen up to $\sim$70 keV.}
\label{erpp}
\end{figure*}

For analysis, we used HEASoft software package (version 6.12).  Calibration database (CALDB) 
files, released on 2012 February 10 (for XIS) and 2011 September 13 (for HXD) by the instrument 
teams are used for data reduction.  The unfiltered event files are reprocessed using `aepipeline' 
package of FTOOLS. These reprocessed cleaned event files are used for further analysis. 
The arrival times of X-ray photons were converted to same at the solar system barycenter
by applying   ``aebarycen'' task of FTOOLS on the reprocessed cleaned event files of
XIS, PIN and GSO. Source light curves and spectra were accumulated from XIS reprocessed 
cleaned event data by selecting a circular region of $3'$ diameter around the central X-ray 
source. The XIS background spectra were extracted from the same event files by selecting
circular regions away from the source position. By using `xisrmfgen' and `xissimarfgen' 
tasks of FTOOLS, the response files and effective area files for each XIS were generated
for spectral fitting. Using reprocessed and cleaned HXD data, light curves and spectra 
were accumulated by using the task ``XSELECT'' of FTOOLS. Simulated background event data
(provided by the instrument teams) were used to estimate  HXD/PIN and HXD/GSO background
for  4U~1909+07 observation. The response files released on 2010 July (for HXD/PIN) and
2010 May (for HXD/GSO) were used for spectral analysis. Additional effective area file, 
released on 2010 May, was used for HXD/GSO.

\section{Results and Discussion}

\subsection{Timing Analysis}

As described above, source light curves with time resolutions of 2 s, 1 s, and
1 s were extracted from barycenter corrected XIS, HXD/PIN and HXD/GSO reprocessed 
event data. The orbital period of the binary system is small (4.4 days) and  
$Suzaku$ observation of the pulsar was spanned over a significant part of the binary 
orbit, it is possible that the X-ray pulsations get smeared. To neutralize the effect 
of orbital motion on the X-ray pulsations, the photon arrival times in each of the
light curves were corrected for the binary motion by using the ephemeris give by
Levine et al. (2004). The orbital motion and barycenter corrected light curves 
were used for timing studies of the pulsar. By applying pulse folding and $\chi^{2}$ 
maximization technique, the pulse period of the pulsar was estimated to be 604.11$\pm$0.14 s 
for XIS-0 and 604.08$\pm$0.13 s for HXD/PIN. The pulse periods of 604.11 s estimated 
from the XIS light-curve was  used to 
generate pulse profiles of the pulsar in 0.4-12 keV, 10-70 keV and 40-200 keV energy 
ranges by using background subtracted XIS, HXD/PIN and HXD/GSO light curves, respectively.
The corresponding pulse profiles are shown in Figure~\ref{pp}. 
From the figure, it is seen that the pulse profiles are strongly energy dependent. In soft 
X-rays (top panel of the figure), the shape of the profile is found to be complex because of 
the presence of dip like features whereas it becomes a single peaked profile in 10-70 keV 
(HXD/PIN) energy range (middle panel). The pulsations are either absent or marginal 
in 40-200 keV (HXD/GSO) energy range (bottom panel). To investigate the evolution of pulse 
profile with energy, several energy resolved light curves were extracted from barycenter 
corrected XIS,  HXD/PIN and HXD/GSO event data. Orbital correction was applied to all the
light curves before generating the pulse profiles. The corresponding energy resolved 
pulse profiles are shown in Figure~\ref{erpp}. The evolution of pulse profile from a 
complex shape at soft X-rays to a single-peaked profile up to $\sim$70 keV can be clearly 
seen in the Figure~\ref{erpp}. It can be seen that 604.11 s pulsation is present in GSO light
curves up to 70 keV energy range beyond which it is absent.

Pulse profiles of the HMXB pulsar 4U~1909+07 is strongly depends on energy.
At low energies ($\leq$10 keV), the shape of the profile is found to be complex because 
of the presence of several dip like structures. These structures disappear at high energies
making the pulse profile single-peaked in 10-70 keV energy range beyond which the pulsations
were absent in the pulsar. Energy dependent dips or dip-like features in the pulse 
profile are seen in many accretion powered X-ray pulsars such as 4U~0115+63 (Tsygankov et al. 
2007), A~0535+262 (Naik et al. 2008),  1A~1118-61 (Maitra et al. 2012 and references therein), 
GRO~J1008-57 (Naik et al. 2011), EXO~2030+375 (Naik et al. 2013) etc. Detailed pulse phase 
resolved spectral analysis of many HMXB pulsars such as GRO~J1008-57 (Naik et al. 2011), 
1A~1118-61 (Devasia et al. 2011a; Maitra et al. 2012), GX~304-1 (Devasia et al. 2011b), 
EXO~2030+375 (Naik et al. 2013) etc.  showed complex pulse profile structure which 
were interpreted as due to the presence of an additional stream of matter at certain 
pulse phases which are phase-locked to the neutron star. Absorption of soft X-ray 
photons by the additional matter in narrow streams causes dips or dip like features in 
the pulse profiles.

\subsection{Spectral Analysis}

The pulse phase averaged spectral analysis of 4U~1909+07 was performed using spectra
accumulated from XIS-0, XIS-1, XIS-3,  PIN and GSO detectors. The corresponding background
spectra and response files were obtained as described above. Spectra from both  FI CCDs 
(XIS-0 and XIS-3) and corresponding background spectra and response files were merged together
by applying the task `addascaspec'. Simultaneous spectral fitting was carried out by using
spectrum, background and response data files of  merged FI CCDs (0.8-10 keV),  XIS-1 (0.8-10 keV),
HXD/PIN (12-70 keV) and HXD/GSO (40-100 keV) with the software package XSPEC v12.7. Because of 
the presence of known artificial structures in XIS energy spectra at around the Si and Au edge, 
data in 1.7-1.9 keV and 2.2-2.4 keV energy ranges were ignored from spectral fitting. XIS spectra 
were re-binned by a factor of 8 from 0.8 keV to 2 keV and by a factor of 6 from 2 keV to 10 keV. 
HXD/PIN spectrum was re-binned by a factor of 2 from 25 keV to 40 keV and by a factor of 6 from 
40 keV to 70 keV. The binning of GSO spectrum was done as suggested by the instrument team. 
All spectral model parameters except the relative instrument normalizations were tied together 
in the spectral fitting. We attempted to fit the broad-band continuum spectrum of
the pulsar with (i) a simple power-law model, (ii) a high energy cut-off power-law model, 
(iii) the NPEX model and (iv) Comptonization continuum model (Sunyaev \& Titarchuk 1985). 
Partial covering absorption component $pcfabs$ was also applied to above continuum 
models in spectral fittings. Apart from these, additional components such as photoelectric 
absorption, Gaussian function for iron emission line, a black-body component for soft X-ray 
excess were needed to describe the continuum spectrum. All these models were reasonably well 
fitted up to $\sim$40 keV beyond which significant anomaly in residuals was noticed in 
the spectral fitting. We tried to obtain a suitable model that describes the 
broad-band spectrum of the pulsar in 0.8-100 keV energy range. 
While fitting XIS, HXD/PIN and HXD/GSO spectra simultaneously, we noticed 
that the relative normalization for HXD/GSO detector was high ($\geq$2) compared
to XIS and HXD/PIN for all above continuum models. The relative normalizations for
HXD/PIN and HXD/GSO should be same or comparable according to the HXD calibration. Apart
from high value of relative instrument normalization for HXD/GSO, the spectrum beyond
100 keV also showed patterns similar to the modeled GSO background spectrum.
As the pulsar is very weak in hard X-rays and there is background-like pattern in 100-150
keV source spectrum after background subtraction, it is likely that the HXD/GSO background is
underestimated for this $Suzaku$ observation. Considering this, 
we did not use the HXD/GSO spectrum in spectral fitting and fitted 
0.8-70 keV spectrum extracted from XIS and HXD/PIN data by using above models.

\begin{figure}
\centering
\includegraphics[height=3.2in, width=2.3in, angle=-90]{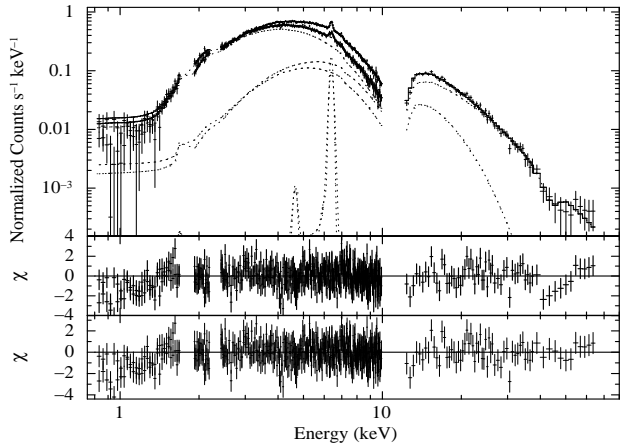}
\caption{Energy spectrum of 4U~1909+07 obtained with  XIS and HXD/PIN detectors of $Suzaku$ 
observation, along with the fitted model comprising  a partial covering power-law continuum 
model along with a black-body component, a narrow iron line emission, and a CRSF. The middle 
and bottom panels show the contributions of the residuals to $\chi^2$ for each energy bin for 
the partial covering power-law continuum model without and with CRSF component in the model, 
respectively.}
\label{spec1}
\end{figure}

\begin{figure}
\centering
\includegraphics[height=3.2in, width=2.3in, angle=-90]{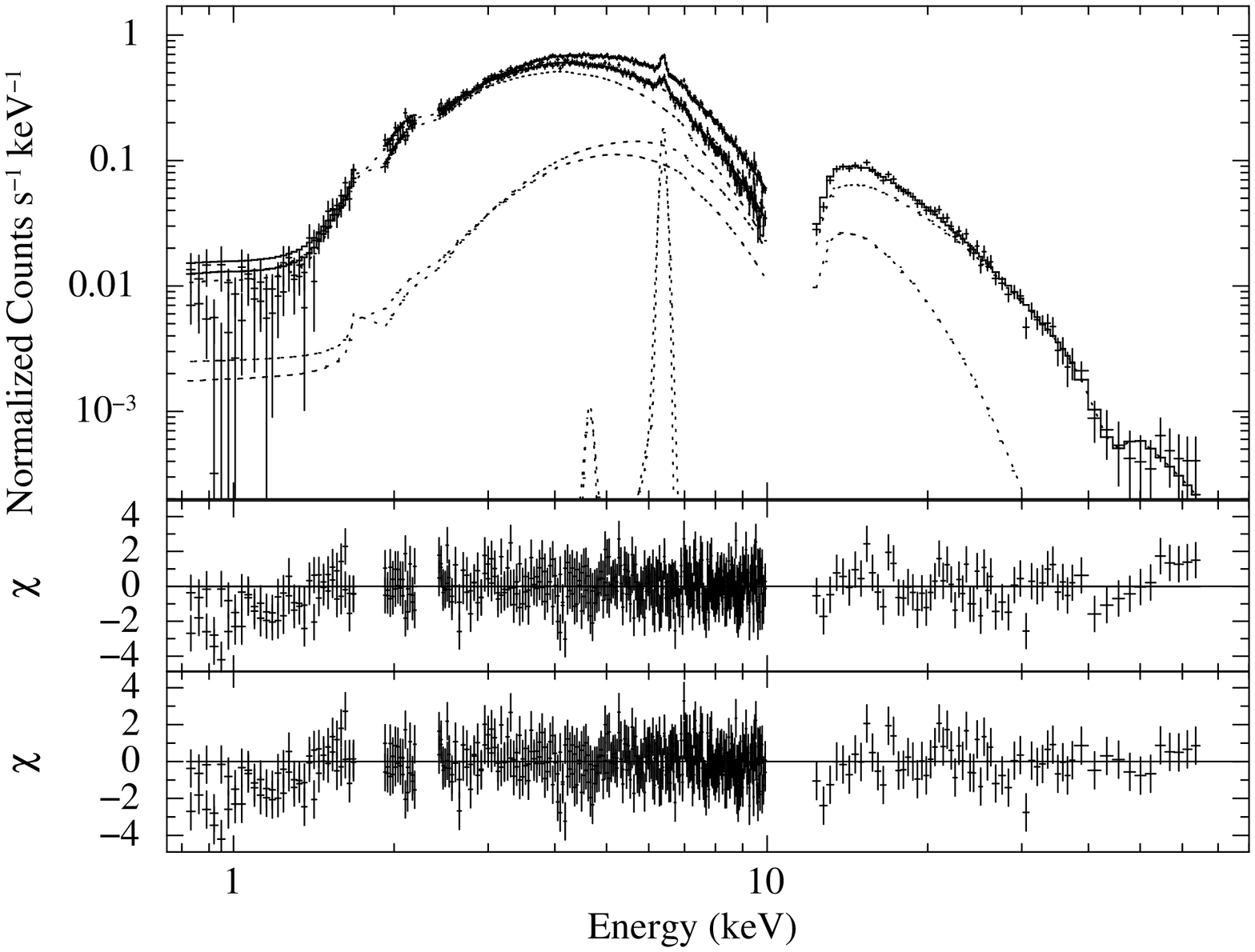}
\caption{Energy spectrum of 4U~1909+07 obtained with XIS and HXD/PIN detectors 
of  $Suzaku$ observation, along with the fitted model comprising a partial covering high energy 
cut-off power-law continuum model along with a black-body component, a narrow iron line emission, 
and a CRSF. The middle and bottom panels show the contributions of the residuals to $\chi^{2}$ for 
each energy bin for the partial covering power-law continuum model without and with CRSF 
component in the model, respectively.}
\label{spec2}
\end{figure}

\begin{figure}
\centering
\includegraphics[height=3.2in, width=2.3in, angle=-90]{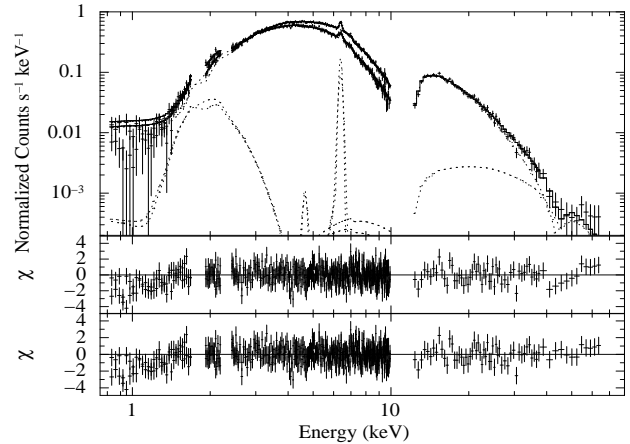}
\caption{Energy spectrum of 4U~1909+07 obtained with XIS and HXD/PIN detectors 
of  $Suzaku$ observation, along with the best-fit model comprising a partial covering NPEX 
continuum model along with a black-body component, a narrow iron line emission, and a CRSF. The middle 
and bottom panels show the contributions of the residuals to $\chi^{2}$ for each energy bin 
for the partial covering NPEX continuum model without and with CRSF component in the model, 
respectively.}
\label{spec3}
\end{figure}

The partial covering power-law continuum model with iron emission 
line and a black-body component of temperature $\sim$3 keV fitted the pulsar spectrum well. 
A black-body component with high temperature as $\sim$3 keV is unusual in case of 
accretion powered X-ray pulsars. The addition of high energy cut-off to the partial covering 
power-law model fitted data well with the black-body temperature of $\sim$0.2 keV. The parameters 
obtained by fitting the partial covering high energy cut-off power-law model to the spectrum are 
in good agreement with Table~2 of F{\"u}rst et al. (2012). However, the partial covering NPEX 
continuum model with iron line and black-body component described the spectrum better yielding an 
acceptable black-body temperature of 0.2 keV. An absorption like feature was seen in pulsar 
spectrum and in the residuals at $\sim$44 keV that allowed to add a CRSF component in 
the spectral model. The addition of CRSF component to above continuum models improved $\chi^2$ 
values for each model (as given in Table~1). The count rate spectra of the pulsar 4U~1909+07 
are shown in Figure~\ref{spec1} (for partial covering power-law model), Figure~\ref{spec2} 
(for partial covering high energy cut-off power-law model) and Figure~\ref{spec3} (for partial 
covering NPEX model) along with the model components (top panels). The middle panels in above three 
figures show the residuals to the fitted models without the CRSF whereas the bottom panels 
show the residuals to the best-fitting model using the CRSF component in the spectral models. The 
presence of an absorption feature at $\sim$44 keV can be clearly seen in the middle panels of all 
figures. 

To test the statistical significance of the $\chi^2$  improvement due to the 
addition of the CRSF component, we performed an F-test. In case of emission line 
features (additive components in $\small{XSPEC}$), the $F$-test routine incorporated 
in $\small{XSPEC}$ package should be the best suited to perform the significance test 
(though care should be taken while using this - see Protassov et al. 2002). However, 
in case of multiplicative components such 
as CRSF, the $F$-test in $\small{XSPEC}$ is inappropriate for performing the null 
hypothesis test. The $F$-test in $\small{XSPEC}$ is based on the assumption that 
the inclusion of the new component does not alter the continuum. 
This is true if the component is added to the continuum model, whereas this is not correct if the 
component is multiplied by the continuum. Therefore, we used a different $F$-test, as described in
Press et al. (2007), to test the statistically significance of the CRSF component in the spectrum
of 4U~1909+07 (see e.g. Orlandini et al. 2012). The $F$-test routine is available in $\small{IDL}$ package 
(named as $mpftest$\footnote{http://www.physics.wisc.edu/$\sim$craigm/idl/down/mpftest.pro}) 
and was used for significance test of multiplicative components such as CRSF or Gaussian 
absorption features (Decesar et al. 2013). The probability of chance improvement (PCI) is 
evaluated for each of the three models used to fit the pulsar spectrum without and with CRSF 
component. The estimated PCI values after addition of CRSF component to the  (i) partial 
covering power-law model with black-body, (ii) partial covering high-energy cutoff 
power-law model with black-body and (iii) partial covering NPEX model with black-body are found 
to be  37\%, 51\% and 47\%, respectively. Considering the high value of PCI for all three models, 
the CRSF is found to be not statistically significant. But the residuals shown in the
middle panels of Figures~\ref{spec1}, \ref{spec2} \& \ref{spec3} clearly show an absorption-like 
feature at $\sim$44 keV. It may be noted that the goodness-of-fit estimator chosen here to 
asses the statistical significance of the CRSF (the $\chi^2$) is not the best suited as it 
does not take into account the ``shape'' of the residuals. Considering the identical 
distribution of residuals around 40 keV (middle panels of Figures~\ref{spec1}, \ref{spec2} 
\& \ref{spec3}, we applied run-test (also called Wald-Wolfowitz test) on the residuals 
obtained from the spectral fitting by using above three continuum models. 
\small{IDL} routine for run-test\footnote{$http://www.astro.washington.edu/docs/idl/cgi-bin/getpro/library07.html?R\_TEST$} was used to derive the null hypothesis of the randomness in 
the residuals of spectral fitting in 37-65 keV energy range. It was found that the number
of data points used for the run-test in 37-65 keV energy range are 12 (6 points below zero
and 6 points above zero), 13 (4 points below zero and 9 points above zero) and 13 (6 points
below zero and 7 points above zero) for partial covering power-law, partial covering high-energy 
cutoff and partial covering NPEX continuum models, respectively. The probability for getting 3 runs 
in above energy range was estimated to be 0.8\%, 0.7\% and 0.5\%, respectively. Marginal
difference in the values of probability is because of the use of different continuum
models that can affect the residuals in the spectral fitting. The computed probability 
of $\leq$1\% for all three continuum models reject the hypothesis of random sampling of 
the detected absorption feature in the pulsar spectrum. Among the three continuum models, 
it is found that the NPEX continuum model with 
black-body, a Gaussian function and a CRSF feature at $\sim$44 keV yields the best fit to 
0.8-70 keV data. The presence of an absorption feature at $\sim$44 keV in the residuals of all 
the models suggest that the CRSF is indeed required in the spectral fitting. The best-fit 
spectral parameters obtained by using three different continuum models along with additional 
components are given in Table~1. It can be seen from the table that addition of a CRSF at $\sim$44 
keV improved the spectral fitting in all cases.

\begin{table*}
\centering
\caption{Best-fit parameters of the phase-averaged spectra for 4U~1909+07
of $Suzaku$ observation with 1$\sigma$ errors. Model-1 : Partial covering
power-law model with black-body and Gaussian components, Model-2 : Partial covering
power-law model with black-body, Gaussian components and cyclotron line. Model-3 : Partial 
covering high energy cut-off power-law model with black-body and Gaussian components, 
Model-4 : Partial covering high energy cut-off power-law model with black-body, 
Gaussian components and cyclotron line. Model-5 : Partial covering NPEX model 
with black-body and Gaussian components, Model-6 : Partial covering NPEX model 
with black-body, Gaussian components and cyclotron line}
\begin{tabular}{lllllll}
\hline
Parameter      		&\multicolumn{6}{c}{Value} 	 \\
                                     &Model-1         &Model-2       &Model-3        &Model-4           &Model-5        &Model-6\\
\hline
N$_{H1}$ (10$^{22}$ atoms cm$^{-2}$) &6.16$\pm$0.17	 &5.95$\pm$0.19  &7.50$\pm$0.21  &7.53$\pm$0.25  &7.09$\pm$0.20   &7.05$\pm$0.19\\
N$_{H2}$ (10$^{22}$ atoms cm$^{-2}$) &9.71$\pm$0.95	 &8.20$\pm$1.28  &32.19$\pm$8.20 &34.01$\pm$7.82 &38.78$\pm$7.89 &39.71$\pm$7.90\\
Covering Fraction                    &0.52$\pm$0.03	 &0.47$\pm$0.05  &0.18$\pm$0.04  &0.18$\pm$0.03	 &0.19$\pm$0.03  &0.20$\pm$0.03\\
$kT_{BB}$ (keV)                 &3.38$\pm$0.21       &2.71$\pm$0.23  &0.19$\pm$0.01  &0.19$\pm$0.01   &0.20$\pm$0.01  &0.20$\pm$0.01\\
Norm.$^a$ of $kT_{BB}$ (10$^{-3}$)           &1.50$\pm$0.14       &1.18$\pm$0.18  &10.59$\pm$3.25 &11.07$\pm$3.10  &5.65$\pm$1.70  &5.27$\pm$1.82\\
Iron line Energy (keV)          &6.39$\pm$0.01       &6.39$\pm$0.01  &6.39$\pm$0.01  &6.39$\pm$0.01      &6.39$\pm$0.01  &6.39$\pm$0.01\\
Iron line width  (keV)          &0.01$\pm$0.01	     &0.01$\pm$0.01  &0.01$\pm$0.01  &0.01$\pm$0.01      &0.01$\pm$0.01  &0.01$\pm$0.01\\
Iron line eq. width (eV)        &73$\pm$3	           &72$\pm$3       &68$\pm$3       &68$\pm$3           &69$\pm$3       &68$\pm$3\\
Power-law index                 &1.91$\pm$0.03	     &1.77$\pm$0.05  &1.43$\pm$0.05  &1.44$\pm$0.04      &1.05$\pm$0.06  &1.01$\pm$0.07\\
Norm.$^b$ of Power-law (10$^{-2}$)    &7.40$\pm$0.51       &5.36$\pm$0.71  &4.16$\pm$0.50  &4.26$\pm$0.52      &                & \\
High energy cut-off (keV)	      &              &               &7.34$\pm$0.28  &7.37$\pm$0.16      &13.98$\pm$1.77 &12.78$\pm$1.58\\
Folding energy  (keV)           &                    &               &23.42$\pm$1.51 &24.23$\pm$1.10     &               &\\
Cyclotron line E$_{a}$ (keV)    &                    &43.75$\pm$1.51 &               &43.10$\pm$2.09     &               &43.85$\pm$1.58\\
Width of cyclotron line	(keV)   & 		     &4.72$\pm$2.58  &               &1.0$^{+2.5}_{-1.0}$    &          &2.04$\pm$2.02\\
Depth of cyclotron line	        &                    &1.25$\pm$0.50  &               &1.01$\pm$0.83     &               &1.57$\pm$0.89\\
Flux$^c$ (in 1-10 keV range)     &1.51$\pm$0.10       &1.51$\pm$0.03   &1.51$\pm$0.18  &1.51$\pm$0.16     &1.51$\pm$0.15  &1.51$\pm$0.14 \\
Flux$^c$ (in 10-70 keV range)    &4.22 $\pm$0.29      &4.14$\pm$0.40   &3.76$\pm$0.42  &3.76$\pm$ 0.38    &3.94$\pm$0.69  &4.01$\pm$0.75 \\
C$_{XIS-03}$/C$_{XIS-1}$/C$_{PIN}$   &1.0/0.94/1.11  &1.0/0.94/1.22  &1.0/0.94/1.18  &1.0/0.94/1.17     &1.0/0.94/1.13  &1.0/0.94/1.13\\
$\chi^2$ (dofs)                      &609 (489)	     &587 (486)     &576 (487)       &573 (484)      &578 (487)     &570 (484)\\
\hline
\end{tabular}
\\
N$_{H1}$ = Equivalent hydrogen column density, N$_{H2}$ = Additional hydrogen column density, 
$^a$ : in units of 10$^{39}$ erg s$^{-1}$ {(d/10 kpc)}$^{-2}$ where $d$ is the distance to the source,
$^b$ : photons keV$^{-1}$cm$^{-2}$s$^{-1}$ at 1 keV, 
$^c$ : Absorption uncorrected flux (in units of \ensuremath{10^{-10}\,  \mathrm{ergs}\,  \mathrm{cm}^{-2}\,\mathrm{s}^{-1}}).

\label{spec_par}
\end{table*}

\begin{figure}
\centering
\includegraphics[height=2.8in, width=3.2in, angle=-90]{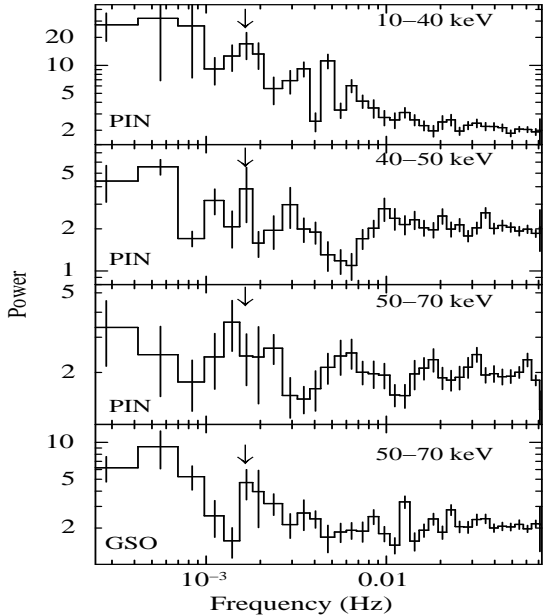}
\caption{The power density spectra of the HMXB pulsar 4U~1909+07 in 10-40 keV (top panel), 
40-50 keV (second panel) and 50-70 keV (third and fourth panels) ranges are shown. Background 
subtracted and orbital corrected HXD/PIN  and HXD/GSO light curves were used to generate power density 
spectra in each energy range. The arrows indicate the spin period of the pulsar. Additional power can be 
seen at $\sim$1.65 mHz (spin frequency of the pulsar) in all panels}.
\label{pds}
\end{figure}

The same $Suzaku$ observation was analyzed, although with the ISIS package,
by F{\"u}rst et al. (2012).  These authors did not use HXD/GSO data because
of possible contamination from the nearby source GRS~1915+105, and
therefore performed their spectral analysis in the 1-40 keV energy range.
Considering the earlier result, the possible CRSF feature at $\sim$44 keV was carefully 
examined by different approaches. As the pulsar is weak in hard X-ray energy ranges, we tried to 
establish that the observed photons in the spectrum beyond $\sim$40 keV are not affected by 
detector energy response uncertainties. We extracted pulsar light curves in 10-40 keV, 40-50 keV 
and 50-70 keV energy ranges from HXD/PIN and HXD/GSO reprocessed event data. 
Corresponding background light curves were also extracted from the simulated background event data 
(as described in the previous section). The background subtracted and orbital corrected light curves 
in above energy ranges were used to generate power density spectra. The energy resolved power 
density spectra of 4U~1909+07, as shown in Figure~\ref{pds}, show additional power (peaks) at $\sim$1.65
mHz (the spin frequency of the pulsar) in 10-40 keV (top panel), 40-50 keV (second panel) and 50-70 keV 
(third and fourth panels) energy ranges. The power density spectra shown in top three panels were
extracted from corresponding light curves obtained from HXD/PIN data where as the one shown in bottom
panel was extracted from HXD/GSO data. The presence of additional power at $\sim$1.65 mHz in the power 
density spectra in 40-50 keV and 50-70 keV confirms that photons beyond 40 keV in the spectrum are not 
associated with any  energy response uncertainties.

\begin{figure}
\centering
\includegraphics[height=3.3in, width=2.in, angle=-90]{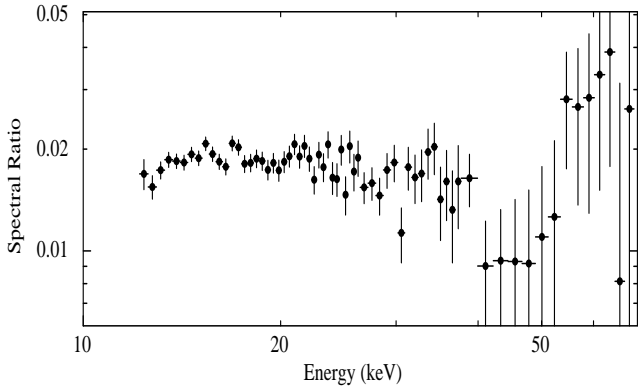}
\caption{Ratio between the background subtracted HXD/PIN spectra of
the HMXB pulsar 4U~1909+07
and Crab Nebula. The absorption feature at $\sim$44 keV is clearly seen.}
\label{cr}
\end{figure}

To establish the absorption like feature seen at $\sim$44 keV in the pulsar spectrum as CRSF,
we evaluated the energy spectra of the pulsar 4U~1909+07 in a model-independent manner. 
We attempted to normalize the pulsar spectrum with that of Crab Nebula, the spectrum of which is a 
featureless power-law with a photon index of $\sim$2.1. The normalization (Crab ratio) has the 
advantage of minimizing the effects due to the detector response and the uncertainties in the 
energy response. To generate Crab ratio, we used a Crab observation with $Suzaku$ (on 2010 April 
5) that is nearest to the observation of the pulsar 4U~1909+07. Data reduction and background 
estimation for  HXD/PIN data of Crab observation were done as described above and the 
Crab ratio was obtained to investigate the presence of the absorption feature in the pulsar 
spectrum. The resulting Crab ratio in 12-70 keV energy range is shown in Figure~\ref{cr}.  The 
presence of the absorption feature at $\sim$44 keV, as was seen in the spectral fitting (middle
panels of Figures~\ref{spec1}, \ref{spec2} \& \ref{spec3}, can be clearly seen in the figure. 
It should be noted here that the shape of this CRSF-like feature in Crab ratio was very similar
to that seen in the middle panels of above figures. In order to evaluate the statistical significance 
of the absorption feature, we performed a run-test on the Crab-ratio data. It was found
that in 37-65 keV energy range in Crab ratio, there are 13 data points with 3 runs in which 7 points are 
above zero and 6 points are below zero. The probability of obtaining 3 runs is only $\sim$1\%. This 
result, together with the results obtained from the analysis on the spectral residuals, 
strongly supports the genuine presence of absorption feature in the Crab ratio at $\sim$44 keV. 
It confirms the presence of CRSF at $\sim$44 keV in the HMXB pulsar 4U~1909+07.

\section{Conclusion}
In this work, we have studied timing and spectral properties of the HMXB pulsar 4U~1909+07 
using data from Suzaku observation. The pulse profiles are strongly depend on energy and evolve 
from complex shape at lower energy to single peaked profile at higher energy. Broad-band 
spectroscopy of the HMXB pulsar 4U~1909+07 in 1-70 keV energy range is reported for the first 
time here. Though the pulsar is very weak in hard X-rays, the high sensitivity
of the HXD onboard $Suzaku$ helped in performing phase-averaged spectroscopy up 
to as high as 70 keV. The energy spectrum of X-ray pulsar is well described by 
partial covering NPEX model with black-body component. For the first time, we report 
the possible detection of CRSF at $\sim$44 keV in this pulsar. Based on this detection, 
the magnetic field strength at the neutron star surface is estimated to be 
$\sim$3.8$\times$10$^{12}$ Gauss. It is seen that the values of CRSF detected in 
accretion powered X-ray pulsars follow a continuum distribution starting from as low 
as $\sim$11 keV for 4U~0115+634 (Nakajima et al. 2006) to as high as $\sim$76 keV 
for GRO~J1008-57 (Yamamoto et al. 2013). Nevertheless a CRSF at $\sim$100 keV 
is reported in LMC~X-4 (La Barbera et al. 2001), it is yet to be confirmed. Considering the 
confirmed values of CRSF in X-ray pulsars, the CRSF detected in the pulsar 4U~1909+07 in 
present work falls in the higher side. It is to be noted that we report the detection of 
CRSF at $\sim$44 keV in the spectrum of 4U~1909+07, obtained from a $\sim$22 ks of exposure 
with HXD. $Suzaku$ observation with long exposure is required to confirm the presence of the 
absorption feature in this pulsar. The LAXPC instrument of the upcoming $Astrosat$ mission will 
also provide a very good opportunity to establish the CRSF feature in the slow HMXB pulsar 
4U~1909+07.

\section*{Acknowledgments}
The authors would like to thank the anonymous referee for his/her constructive comments and 
suggestions that improved the contents of the paper. The research work at Physical Research 
Laboratory is funded by the Department of Space, Government of India. The authors would like 
to thank all the members of Suzaku for their contributions in the instrument preparation, 
spacecraft operation, software development, and in-orbit instrumental calibration. This 
research has made use of data obtained through HEASARC Online Service, provided by NASA/GSFC, 
in support of NASA High Energy Astrophysics Programs.

\end{document}